\documentclass[aps,prl,reprint,superscriptaddress]{revtex4-1}

\usepackage{graphicx}% Include figure files
\usepackage{dcolumn}% Align table columns on decimal point
\usepackage{bm}
\usepackage{amsfonts}
\usepackage{amssymb}
\usepackage{amsmath}
\usepackage{graphicx}
\usepackage{color}
\usepackage{array}
\usepackage{dcolumn} % Align table columns on decimal point
\usepackage{bm} % bold math
\usepackage{amsthm}
\usepackage{latexsym}
\usepackage{float}
\usepackage{tabularx}
\usepackage{lineno}

\begin{document}

\title{Microresonator soliton dual-comb imaging}

\author{Chengying Bao}
\altaffiliation{These authors contributed equally to this work}
\affiliation{T. J. Watson Laboratory of Applied Physics, California Institute of Technology, Pasadena, California 91125, USA}

\author{Myoung-Gyun Suh}
\altaffiliation{These authors contributed equally to this work}
\affiliation{T. J. Watson Laboratory of Applied Physics, California Institute of Technology, Pasadena, California 91125, USA}
\affiliation{NTT Physics and Information Laboratory, 1950 University Ave., East Palo Alto, California 94303, USA}

\author{Kerry Vahala}
\email{vahala@caltech.edu}
\affiliation{T. J. Watson Laboratory of Applied Physics, California Institute of Technology, Pasadena, California 91125, USA}

\begin{abstract}
Fast-responding detector arrays are commonly used for imaging rapidly-changing scenes. Besides array detectors, a single-pixel detector combined with a broadband optical spectrum can also be used for rapid imaging by mapping the spectrum into a spatial coordinate grid and then rapidly measuring the spectrum. Here, optical frequency combs generated from high-$Q$ silica microresonators are used to implement this method. The microcomb is dispersed in two spatial dimensions to measure a test target. The target-encoded spectrum is then measured by multi-heterodyne beating with another microcomb having a slightly different repetition rate, enabling an imaging frame rate up to 200 kHz and fillrates as high as 48 MegaPixels/s. The system is used to monitor the flow of microparticles in a fluid cell.  Microcombs in combination with a monolithic waveguide grating array imager could greatly magnify these results by combining the spatial parallelism of detector arrays with spectral parallelism of optics.
\end{abstract}

\maketitle
%\newcommand{\ts}{\textsuperscript}
%\newcommand{\tsb}{\textsubscript}
%%%%%%%%%%%%%%%%%%%%%%%% Main Text %%%%%%%%%%%%%%%%%%%%%%%
%\begin{document}
%\maketitle
\section{Introduction}
\begin{figure*}[t]
\begin{centering}
\includegraphics[width=0.9\linewidth]{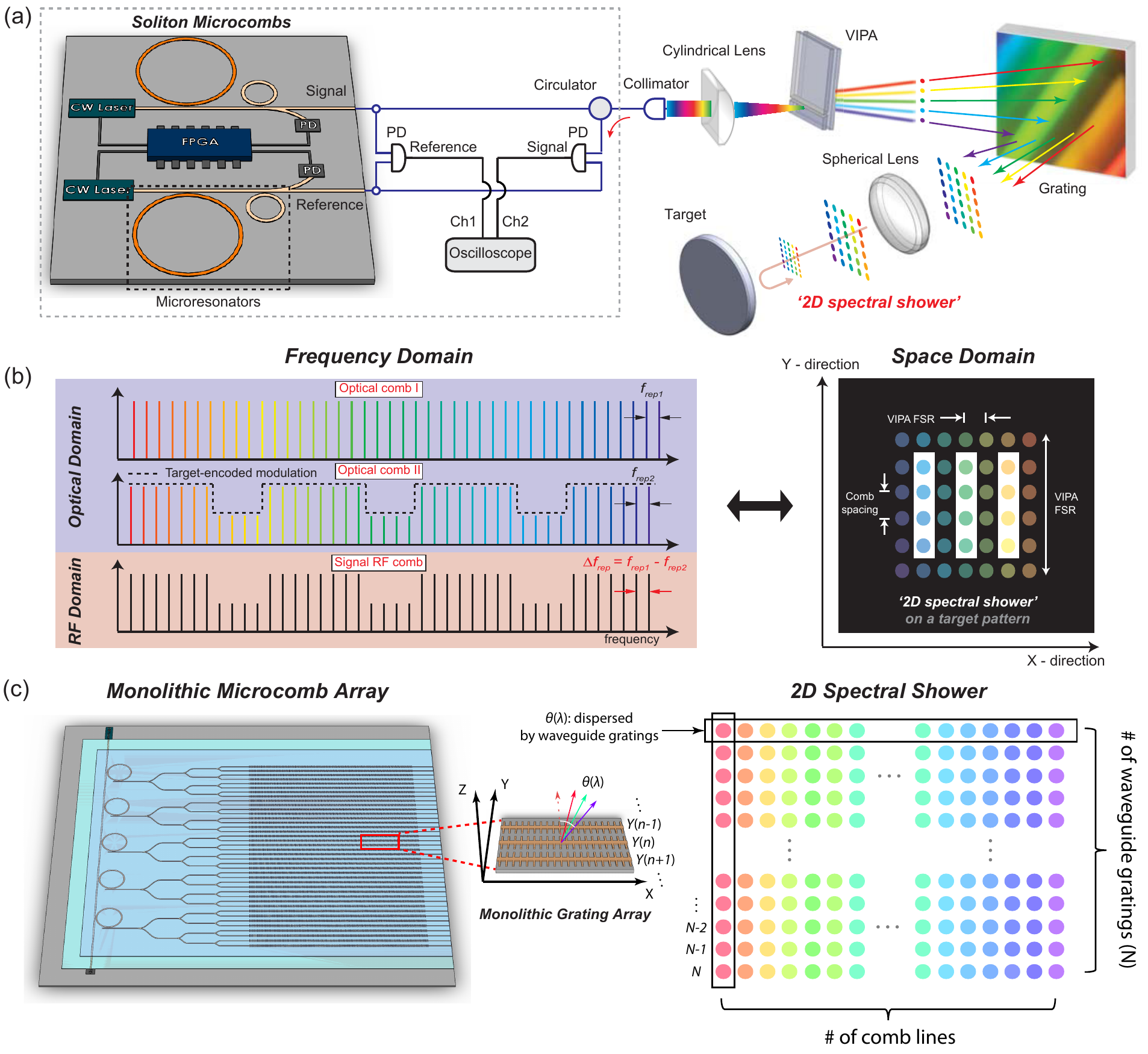}
%\captionsetup{singlelinecheck=no, justification = RaggedRight}
\caption{{\bf Dual-comb imaging using microresonator solitons.} (a) A conceptual diagram showing the operational principle for spectral-spatial-mapping and dual-microcomb imaging. Two soliton microcombs (signal and reference) having slightly different repetition rates are generated using two on-chip microresonators. A 2D disperser (VIPA+grating) maps frequencies from the signal microcomb into a 2D grid of spatial locations (spectral shower) that are reflected by a target. The reflected signal spectrum is measured by multi-heterodyne detection with the reference microcomb. The chip is shown with small (high rate) and larger (low rate) comb pairs in both the signal and reference arms. These can enable different operational modes for the imaging system. %\textcolor{red}{ A larger microresonator,  low-repetition-rate comb (A) and smaller microresonator, high-repetition-rate comb (B) are shown on the chip. Although shown in a tandem configuration, the microcombs could in principle operate on separate waveguides to create a multiplexed array.}  %(b) A photograph showing the top view of the two types of silica microresonators used in the experiment. Free spectral ranges (FSRs) are indicated. %A schematic cross-sectional view of the silica wedge microresonators with the spatial mode intensity indicated (right). (c) Measurement targets used include a USAF 1951 test target and also microparticles within a flow-cell. \textcolor{red}{(d) A pair of high repetition rate combs can be used for high frame rate imaging but with a relatively low resolution (comb B in panel a) while a pair of low repetition rate combs can be used to increase the resolution at a relatively low frame rate (comb A in panel a)}.
(b) Dual-comb imaging proceeds by illuminating the target (right panel) with the 2D spectral shower formed as shown in panel a. As shown in the left panel, the target reflection amplitude is encoded onto the signal comb (Optical comb \textrm{II}). The signal comb is then heterodyned with the reference comb (Optical comb \textrm{I}) to generate the RF comb. $f_\text{{rep1}}$, $f_\text{{rep2}}$, and $\Delta f_\text{{rep}}$ are the frequency line spacing of the reference comb, the signal comb, and the signal RF comb. (c) Dual comb imaging concept based on integrated waveguide grating antennas. Microcomb outputs are divided into multiple waveguides that drive the grating antennas. Comb light is   dispersed by a corresponding waveguide grating antenna (eliminates VIPA and grating) to create one imaging dimension in the spectral shower (right). The second imaging dimension is provided by the spatial location of each grating antenna. This approach combines spectral parallelism of photonics with spatial parallelism of detector arrays to greatly magnify performance. A single (shared) pump is shown, but the microcombs could also be individually pumped so as to create frequency combs that are spectrally displaced.}
\label{Fig1Scheme}
\end{centering}
\end{figure*}

The development of the rapid-frame-capture detector array sensors based on CCD (charge-coupled device) and CMOS (complementary metal oxide semiconductor) technology has revolutionized imaging \cite{Shimadzu2012}.  %Also, by combining CCD/CMOS with streak cameras \cite{StreakCamera}, frame rates of 100 billion per second are possible \cite{Wang_Nature2014}. However, streak cameras are complicated and expensive systems. Moreover, high speed CCD/CMOS-based imaging encounters challenges such as compromised sensitivity at shorter exposure times, heat concentration, and on-chip storage memory and electronic readout speed \cite{Shimadzu2012,Armstrong_Sensor2009}.
Recently, there has also been interest in new methods that leverage the massive bandwidth of optical signals to perform imaging using a single pixel detector. One such approach uses the broadband spectrum of ultrashort optical pulses \cite{Jalali_Nature2009} and works by mapping different optical frequencies of the broadband spectrum into distinct spatial locations using spatial dispersers such as demonstrated in the technique of femtosecond pulse shaping \cite{Weiner_RSI2000}. To create a 2-dimensional (2D) map, a conventional grating disperses the spectrum in one spatial dimension, while a virtually imaged phase array (VIPA) disperses light into the other spatial dimension.  As shown in Fig. \ref{Fig1Scheme}(a), the grating and the VIPA create a `2D spectral shower' in which distinct optical frequencies have a one-to-one (spectral-spatial) correspondence with spatial coordinates in 2-dimensions \cite{Weiner_OE2008TwoD,Diddams_Nature2007,Jalali_Nature2009,Yasui_Optica2018}. To recover the image, the spectrum can be measured by the time-stretch method, which converts the spectrally-encoded spatial information into a temporal waveform measured on the single-pixel photodetector \cite{Jalali_Nature2009}. This approach measures the image on a shot-by-shot basis and 6 MHz frame rates have been demonstrated \cite{Jalali_Nature2009}.

An alternative image recovery technique based on two frequency-combs has also been recently demonstrated  \cite{Yasui_Optica2018}. This approach, termed here dual-comb imaging, parallels the technique of dual-comb spectroscopy \cite{Newbury_Optica2016,Picque_NP2019} by converting an optical spectrum into a radio-frequency (RF) electrical signal. In effect the method maps the optical signal comb with target information into these radio-frequency components (see Fig. \ref{Fig1Scheme}(b)). If the signal and reference comb are phase locked then both amplitude and phase information about the target can be retrieved, enabling acquisition of 3 dimensional information \cite{Yasui_Optica2018}. Line-scan spectral-spatial imaging using dual frequency combs has also been recently reported \cite{Zeng_OL2018,ZhangXL_OL2018,Yasui_JQE2019}. To generate broadband optical pulses for imaging, table-top mode-locked lasers have so far been used. A recent advance in optical pulse and frequency comb generation is based on dissipative Kerr soliton mode locking in optical microcavities \cite{Kippenberg_NP2014,Vahala_Optica2015,Gaeta_Ol2016Thermal,Weiner_OE2016,Kippenberg_review2018,Gaeta_NP2019Review}. The devices provide high repetition rate soliton streams and their associated optical frequency combs feature smooth spectral envelopes. These miniature frequency combs or microcombs \cite{Kippenberg_Science2011} are considered a possible way to dramatically reduce the form factor of conventional frequency comb systems. Accordingly, they are being studied for several applications including dual-comb spectroscopy \cite{Vahala_Science2016,Lipson_SA2018}, ranging \cite{Vahala_Science2018Range,Kippenberg_Science2018Range},
optical communications \cite{Kippenberg_Nature2017Commun}, optical frequency synthesis \cite{Diddams_arXiv2017Synthesis}, and exoplant detection in astronomy \cite{obrzud2019microphotonic,suh2019searching}.

%In the method, a signal comb (repetition rate $f_\text{rep2}$) \textcolor{red}{It is a bit strange that we define $f_\text{rep2}$ before $f_\text{rep1}$. I understand it connects to Fig. 1, but readers may need sometime to get it. We have more similar definition about $f_\text{rep}$ in the Discussion section, where I also commented. How about simply skip mentioning $f_\text{rep1, 2}$ here and when we mentioning $\Delta f_\text{rep}$ we can connect to Fig. 1} is used to create a 2D spectral shower of comb lines that are projected to the target. The reflected lines are collected and heterodyned (on a single-pixel detector) with comb lines from a reference comb (repetition rate $f_\text{rep1}$). The resulting line-by-line interference of the signal and reference combs creates a set of radio frequency components (frequency separation $\Delta f_\text{rep} = f_\text{rep1} - f_\text{rep2}$) in the photocurrent.

The application of microcombs to the dual comb imaging method is considered here. These devices offer a system-on-a-chip architecture that eliminates fiber optics (i.e., that required for the time-stretch image recovery method and to generate mode locked pulses). A fully integrated platform that avoids the free space grating and VIPA elements is also possible (see Fig. \ref{Fig1Scheme}(c)). This work explores soliton microcomb dual-comb imaging by measuring a USAF1951 test target and by monitoring microparticles in a flow-cell. An important feature of microcombs is their very high repetition rates as compared to conventional combs (typically microwave to THz rates as compared to radio frequency rates). The impact of such high repetition rates on future dual-comb imaging system performance is also considered.

\begin{figure*}[t]
\begin{centering}
\includegraphics[width=0.95\linewidth]{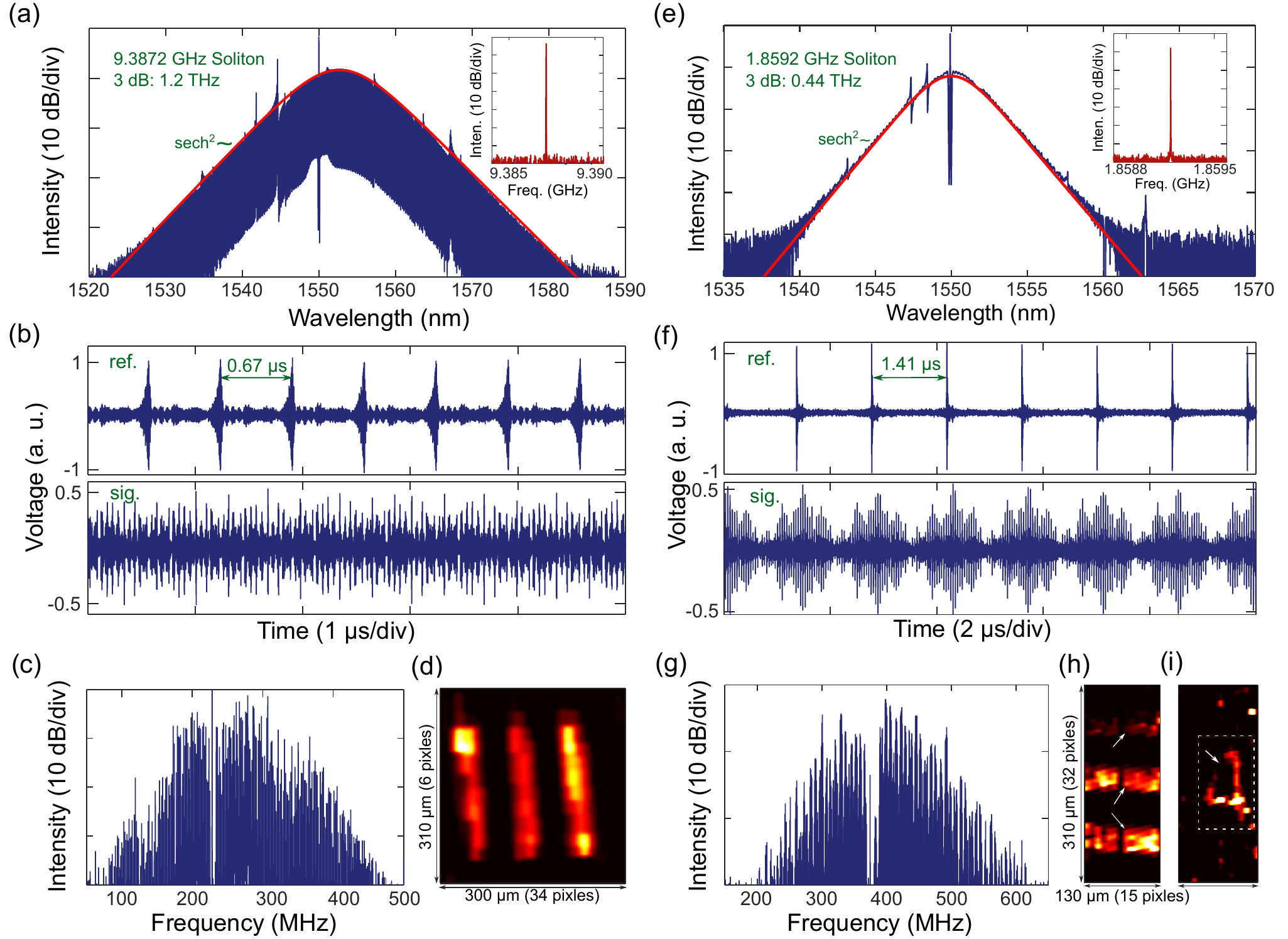}
%\captionsetup{singlelinecheck=no, justification = RaggedRight}
\caption{{ \bf Dual-microcomb imaging of static targets.} (a) Typical optical spectrum of the 9.39 GHz soliton microcombs showing sech$^2$ spectral envelope fit (red) with 3 dB bandwidth of 1.2 THz. The inset is the electrical spectrum of the photodetected soliton pulse stream and gives the repetition rate. (b) An example of the measured interferogram in a 5 $\mu$s time window from the reference arm and the signal arm (see Fig 1(a)). (c) RF spectrum of the 5 $\mu$s signal interferogram in panel b. (d) Image of three vertical bars constructed from the measured interferogram in panel b. (e) Typical optical spectrum of the 1.86 GHz soliton microcombs showing sech$^2$ spectral envelope fit (red) with 3 dB bandwidth of 0.44 THz. A notch near the spectral maximum is produced by narrow-band filtering of the optical pump.  The inset is the electrical spectrum of the photodetected soliton pulse stream. (f) Examples of the recorded interferograms for reference and signal arms using the 1.86 GHz microcombs. (g) RF spectrum of the signal interferogram in panel f. The spectral hole around 380 MHz results from the notch filter used to suppress the optical pump.
(h) Image of 3 horizontal bars constructed from the measured 10 $\mu$s interferogram in panel e. (i) Image of number `4' on the USAF target produced using the 1.86 GHz microcombs. The dark discontinuities shown by the arrows in panels (h, i) result from the spectral notch produced by filtering the optical pump (see panel e). As an aside, a similar discontinuity is located within a dark region of the image in panel d and is therefore not visible.}
\label{Fig3DualComb}
\end{centering}
\end{figure*}

\begin{figure*}[t]
\begin{centering}
\includegraphics[width=0.95\linewidth]{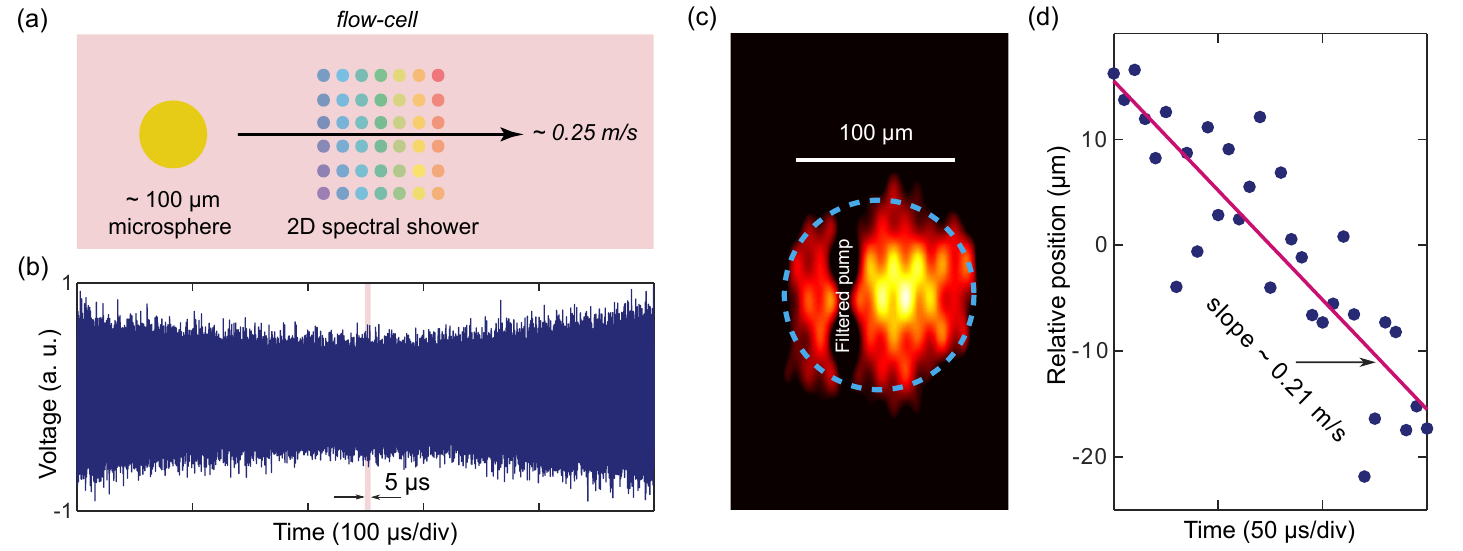}
%\captionsetup{singlelinecheck=no, justification = RaggedRight}
\caption{{ \bf Monitoring flowing particles.} (a) An illustration of the microparticle monitoring experiment. Microparticles are suspended in water and flow inside the cell. When a particle passes through the 2D spectral shower, the particle can be imaged using the dual-comb interferogram. (b) Measured interferogram shows varying amplitude when the microparticle flows through the 2D spectral shower. (c) A snapshot of the measured microparticle, which is constructed from a 5 $\mu$s duration interferogram (shaded bar in panel b). The dashed circle suggests the microparticle size ($\sim$100 $\mu$m). The dark vertical band results from filtering of comb lines around the pump. (d) Center position of the microparticle plotted versus time. A linear fit gives a flow velocity of 0.21 m/s in reasonable agreement with the set water flow velocity of 0.25 m/s.}
\label{Flow}
\end{centering}
\end{figure*}

%\noindent \textbf{Results}
\section{Experimental setup}
High-$Q$ silica-on-silicon wedge microresonators \cite{Vahala_NP2012Wedge} are used to generate the dual soliton streams. The devices feature repetition rates of 1.86 GHz and 9.39 GHz \cite{Vahala_Optica2018GHz}. %Images of the resonators are given in Fig. \ref{Fig1Scheme}(b) and
Details on methods used to trigger and stabilize the soliton microcombs are presented elsewhere \cite{Vahala_Optica2015}. The microcombs are coupled directly to optical fibers using tapered-fiber couplers \cite{Cai2000,Spillane2003}. The signal comb and reference comb are conveyed along the optical train as shown in Fig. \ref{Fig1Scheme}(a). The tapered-fiber couplers can be replaced by integrated waveguides \cite{Vahala_NP2018}. Moreover, fully integrated soliton microcombs with an on-chip pump have also been reported \cite{stern2018battery}.

The VIPA and grating act together to create the 2D spectral shower with the VIPA dispersing the spectrum along the vertical direction and the grating dispersing the spectrum along the horizontal direction. More specifically, the VIPA disperses light only within its free-spectral-range (FSR) which means that optical frequencies $\nu$ and $\nu+mf_\text{VIPA}$ (where $m$ is an integer and $f_\text{VIPA}$ is the VIPA FSR) will overlap in space. By adding a grating, frequencies $\nu$ and $\nu+mf_\text{VIPA}$ can be further dispersed to create the 2D spectral shower as illustrated in Figs. \ref{Fig1Scheme}(a, b). The VIPA used in our experiment has an FSR of 60 GHz (LightMachinery) and this limits pixel count along the vertical direction to 6 pixels (9.39 GHz microcomb) and 32 pixels (1.86 GHz microcomb). Analysis of more optimal designs is provided in the Discussion. The spectral shower is reflected by the object, coupled back into the fiber for return to a photodetector where it is heterodyned with the reference microcomb. Figure \ref{Fig1Scheme}(b) illustrates how the image reflection amplitude is transferred from the spectral shower to the signal RF spectrum produced by dual-comb heterodyne.

Also shown in Fig. \ref{Fig1Scheme}(a) are a collimator and a cylindrical lens (focal length of 150 mm) that focuses the collimated comb onto the VIPA. Additionally, the 2D spectral shower is focused onto the target by a spherical lens (focal length of 30 mm). The targets are placed at the focal plane of the spherical lens and aligned to provide maximum reflection coupled into the fiber. Because the dual-comb measurement can resolve single comb lines, the spatial resolution is set by the imaging system only. This is in contrast to the time-stretch method where spatial resolution can also be limited by the ability to resolve the frequency components \cite{Jalali_Nature2009}. The spot diameter of a dispersed comb line is $\sim$15 $\mu$m in the current setup and a finer spot size can be achieved by expanding the beam size before the focusing spherical lens.

\section{Imaging a static target}
To demonstrate this approach, a USAF 1951 test target (negative) is imaged. %Target patterns within Group 3 Element 4 of the target were illuminated.
In a first measurement, two independent free-running silica microcombs with repetition rates close to 9.39 GHz are used. Dual combs can also be generated from a single microresonator \cite{Vahala_NP2017Counter,Gaeta_OL2018CP,Kippenberg_NP2018CP} which can result in strong common noise suppression but it also limits the freedom of choosing the repetition rate difference. Therefore, two independent microresonators were used. The spectrum of one of the microcombs is shown in Fig. \ref{Fig3DualComb}(a) and features a sech$^2$-shaped spectral envelope and 3 dB bandwidth of 1.2 THz. The spectral spurs in the spectrum result from the dispersive wave emission induced by avoided-mode-crossings, which also assists single soliton generation \cite{Weiner_Optica2017}. The detected electrical spectrum for this comb is shown in the inset of Fig. \ref{Fig3DualComb}(a). The other microcomb has a similar spectrum, but its repetition rate differs from the first microcomb by $\Delta f_{\text{rep}}\sim$1.5 MHz. The repetition rate difference was chosen to maintain the interferogram bandwidth within the 1 GHz bandwidth of the digitizer used in the experiment. The close matching in the selected repetition rates is possible by good microfabrication control of the resonator diameters using a common mask size and calibration of etch rates \cite{Vahala_NP2012Wedge}. In Fig. \ref{Fig3DualComb}(b), typical examples of the heterodyned dual-comb interferograms measured over a 5 $\mu$s interval from the reference arm (upper panel) and signal arm (lower panel) are shown. While the reference interferogram contains a readily identifiable periodic signal resulting from the difference in repetition rates of the signal and reference microcombs, the signal interferogram contains complex structure associated with the image.

To construct an image, an RF spectrum is first calculated by taking the fast Fourier transform (FFT) of the signal interferogram produced by illuminating a patterned region of the target (Fig. \ref{Fig3DualComb}(c)). Even though free running microcombs were used, the signal-to-noise-ratio (SNR) is at least 10 dB over most of the spectrum and exceeds 30 dB over a substantial fraction of the spectrum. Furthermore, the RF spectrum is compressed into a bandwidth less than 400 MHz, a rate that is much smaller than, for example, that required in pulsed imaging work \cite{Jalali_Nature2009}. The signal spectrum is then normalized using the FFT of the reference interferogram. Following this calculation, the same procedure is applied except using a non-patterned (uniform) region of the target. Finally, this non-patterned RF spectrum is used to normalize the RF spectrum of the patterned region,
%The spectral lines with amplitudes larger than 0 dB might result from non-uniformity of the target (for example glints). These amplitudes, however, have a negligible affect on the resulting image because they represent small pixel values in the image (inverse reflectivity is used for reconstruction as a negative target is used).
and the resulting normalized spectrum is sorted into the 2D image matrix with each column containing one VIPA FSR. An example of the constructed image is shown in Fig. \ref{Fig3DualComb}(d). The bars appear slightly tilted because the 2D spectral shower generated by the VIPA and grating is usually titled \cite{Diddams_Nature2007,Weiner_OE2008TwoD}.

The low pixel number ($\sim$6$\times$34) limits the resolution of the image, especially along the vertical direction which is set by the combination of the 9.39 GHz microcombs with the fixed VIPA FSR. To illustrate possible improvement in the vertical direction, two 1.86 GHz soliton microcombs ($\Delta f_{\text{rep}}$=0.7 MHz) were also tested in the imaging setup (see Fig. \ref{Fig3DualComb}(e) for the optical and electrical spectra of one of the microcombs). The pixel number using the 1.86 GHz microcombs is $\sim$15$\times$32. The field of view (shown in Fig. \ref{Fig3DualComb}) is smaller than the 9.39 GHz microcombs due to the narrower comb bandwidth. Reference and signal dual-comb interferograms can be obtained as before (Fig. \ref{Fig3DualComb}(f)) with a corresponding signal FFT (Fig. \ref{Fig3DualComb}(g)). Fig. \ref{Fig3DualComb}(h, i) show the resulting images of three horizontal bars and the number `4' (test target), respectively. Figs. \ref{Fig3DualComb}(h, i) are recorded within a time interval of 10 $\mu$s.

As an aside and as noted earlier, locking of the signal and reference combs can allow for phase retrieval enabling 3 dimensional imaging similar to ref. \cite{Yasui_Optica2018}. Such locking could also leverage the mutual locking of counter-propagating solitons \cite{Vahala_NP2017Counter}, however, this will lower the frame rate due to the relatively small repetition rate difference in counter-propagating solitons. Also, even though the 9.39 GHz microcombs and 1.86 GHz microcombs were generated from different chips in the current experiments, multiple comb pairs can be integrated on a single chip to enable agile switching between different operating modes (see Fig. \ref{Fig1Scheme}(a) for the concept).

\medskip

\section{Monitoring a flowing microparticle}
%The frame rate of the heterodyne image construction approach is determined by the requirement to resolve the RF signal comb. Accordingly, because the electrical comb lines have a frequency separation equal to the difference in the repetition rates of the signal and reference soliton microcombs, this difference in repetition rates sets an upper bound to the frame rate and can be quite high when using soliton microcombs (see Discussion).

To demonstrate measurement of a rapidly-changing scene, two 9.39 GHz soliton microcombs are used to monitor a microparticle moving in a high-speed flow-cell (Fig. \ref{Flow}(a)). For this purpose a $\sim$ 0.25 m/s laminar flow cell (cross section 0.2 mm $\times$ 8 mm) was set up and microparticles with diameter of $\sim$ 100 $\mu$m were suspended in water to flow through the cell. To improve signal-to-noise a mirror was placed behind the cell. When a microparticle passes through the spectral shower, it modulates lines in the spectral shower, which results in the amplitude varying interferogram shown in Fig. \ref{Flow}(b). An image of a recorded flowing microparticle is constructed from the 5 $\mu$s portion of the interferogram (shaded bar) in Fig. \ref{Flow}(b) and is shown in Fig. \ref{Flow}(c). The size of the reconstructed microparticle (dashed circle in figure) is consistent with the particle's actual size. The particles are not well resolved on account of the limited pixel number. To measure the motion of the microparticle, its center position is plotted versus time in Fig. \ref{Fig3DualComb}(d). A linear fit gives a flow velocity of 0.21 m/s, consistent with the flow-cell set-point velocity of 0.25 m/s. A 200 kHz frame-rate was used in this measurement and $\sim$7 frames of the heterodyne interferogram were averaged to produce the image. A comparison of frame rate, pixel count and fillrate for this result versus work using fiber lasers \cite{Yasui_Optica2018} is provided in Table 1.

\medskip

\section{Discussion and outlook}
\subsection{Repetition rate and imaging performance}

Comb repetition rate affects pixel count, fillrate and frame rate in the dual comb imaging system. Pixel count determines target resolution and is equal to the number of comb lines,
\begin{equation}
\label{Pixelcount}
M_1 \times M_2 = B / f_{\text{rep}}
\end{equation}
where $M_1$ and $M_2$ are the pixel count along the two axes of the spectral shower, $B$ is the comb bandwidth and $f_{\text{rep}}$ is the comb repetition rate.

Beyond pixel count, the image processing rate is also important. To better understand the constraints inherent in the dual comb approach, note that the $M_1 \times M_2$ comb line pixels, once mapped into the radio frequency domain, must fit within a radio frequency bandwidth that is less than $f_\text{rep}/2$. If this condition is not satisfied, then the optical to radio-frequency mapping would result in spectral folding of comb components \cite{Newbury_Optica2016}.
This constraint gives $M_1 \times M_2 \times \Delta f_{\text{rep}} < f_{\text{rep}}/2$. The rate $\Delta f_{\text{rep}}$ also sets the maximum frame rate of the imaging system. This can be understood by considering the interference of the the signal and reference combs in the time domain where their different repetition rates causes the combs to strobe one another on the photodetector. Each strobed signal in the detected current contains the complete image information so that the strobing rate (i.e., $\Delta f_{\text{rep}}$) is the maximum image frame rate. In practice, several strobe periods ($C$ periods) must be averaged to improve the signal-to-noise so that the practical frame rate is $f_\text{frame} = \Delta f_{\text{rep}} / C$. As a result of this relationship, the number of pixels that can be detected per second by the dual comb imaging system is given by,
\begin{equation}
\label{Fillrate}
F_1 \equiv M_1 \times M_2 \times f_{\text{frame} } = f_{\text{rep}}/ (2 C)
\end{equation}
where $F_1$ is the fillrate (pixels per second) of a VIPA+grating imager (Fig. \ref{Fig1Scheme}(a)) or a single waveguide antenna in Fig. \ref{Fig1Scheme}(c). Fillrate is widely used to characterize video cards, but here it is used to assess the combined space and time resolution of the dual-comb imaging system. Finally, it is also possible to eliminate $M_1 \times M_2$ in the above expression to arrive at,
\begin{equation}
\label{Framerate}
f_{\text{frame}} < f_{\text{rep}}^2 / (2 B C)
\end{equation}

In summary, Eqns. (1), (2) and (3) show that even while the high repetition rates of microcombs degrade space resolution by reduction of pixel count, they simultaneously improve the frame rates and fillrate of the dual-comb imaging system. Indeed, the fillrates of the current demonstrations are higher than fiber laser based systems (see Table \ref{Design}). Moreover, as discussed in the next section, it is possible to restore pixel count by combining the spatial parallelism of conventional detector arrays with the spectral parallelism of single-pixel photonic imaging as illustrated in Fig. \ref{Fig1Scheme}(c).

\subsection{Design comparisons}

For the current VIPA+grating based system, the assignment of spectral shower pixels to horizontal and vertical axes is controlled through the VIPA FSR ($f_{\text{VIPA}}$) such that $M_1 = f_{\text{VIPA}}/f_{\text{rep}} $ and $M_2 = B / f_{\text{VIPA}} $. %For a square array ($N = M$) one must have $f_{\text{VIPA}}^2 = B f_{\text{rep}}$.
As examples, a 25 THz comb bandwidth would enable a horizontal pixel number of 50 using a 500 GHz VIPA (custom design available at LightMachinery). For the vertical direction, the pixel number could be increased to over 50 using 10 GHz repetition rate microcombs. This configuration would provide 2500 pixels at a fillrate of 1 GigaPixels/s (see Table \ref{Design}).

\begin{table*}[hbtp]
\centering
\caption{Towards optimal design of dual microcomb imaging systems. $^a$ This is the comb bandwidth that is available for the image retrieval and can be greater than the 3 dB bandwidth. $^b$ The 1.2 kHz frame rate for ref. \cite{Yasui_Optica2018} uses a single period of the interferogram which could limit the image quality. Another frame rate (12 Hz) averages 100 periods and was also reported in ref. \cite{Yasui_Optica2018}. $^c$ In the current work, averaging over 5 periods of the interferogram is assumed. $^d$ The array consists of 10 signal microcombs with a 10:1 waveguide grating antenna fanout.}
\begin{tabular}{ccccccc}
\hline
~~ & VIPA FSR &  Comb bandwidth$^a$ & $\Delta f_\text{rep}$ & Frame rate & Pixels &  Fillrate \\
\hline
100 MHz fiber lasers \cite{Yasui_Optica2018} & 15 GHz & 1.2 THz & 1.2 kHz &  1.2 kHz (12 Hz)$^b$ & 12,382 & 15 (0.15) MegaPixels/s \\
9.4 GHz microcombs & 60 GHz & 1.8 THz & 1.5 MHz & 200 kHz & 204 & 40 MegaPixels/s \\
1.9 GHz microcombs & 60 GHz & 0.9 THz & 0.7 MHz & 100 kHz & 480 & 48 MegaPixels/s \\
Possible 10 GHz microcombs & 500 GHz & 25 THz & 2 MHz & 400 kHz$^c$ & 2,500 & 1 GigaPixels/s \\
100 GHz microcomb array $^d$ & N/A & 10 THz & 500 MHz & 100 MHz$^c$ & 10,000 & 1 TeraPixels/s \\
\hline
\end{tabular}
  \label{Design}
\end{table*}

%We also note that the following expression results from the need to avoid spectral folding of the RF comb lines: $M_1 \times M_2 < f_{\text{rep}}/(2 \Delta f_{\text{rep}})$ \cite{Newbury_Optica2016}. Moreover, $\Delta f_{\text{rep}}$ is related to the frame rate by $\Delta f_{\text{rep}} = C f_{\text{frame}}$ where $C$ is the number of interferogram periods that must be intregrated to obtain a desired SNR ($C \sim 7$ in the present work). As a result of these two relations, the following expression results: $F_1 \equiv M_1 \times M_2 \times f_{\text{frame} } < f_{\text{rep}}/ (2 C) $ where $F_1$ is the fillrate (pixels per second) of the imaging system. As a result of their extremely high repetition rates, microcombs can therefore improve the fillrate of dual-comb imaging systems. Indeed, the fillrates of the current demonstrations are higher than fiber laser based systems. By further optimization of the dual-microcomb imaging system, a fillrate up to 1 GigaPixels/s is possible (see Table \ref{Design}). It is also possible to eliminate $M_1 \times M_2$ in the above expression to arrive at $ f_{\text{frame}} < f_{\text{rep}}^2 / (2 B C)$, which illustrates the importance of higher repetition rate to achieve higher frame rate.

It is also possible to eliminate the discrete VIPA and grating components shown in Fig. \ref{Fig1Scheme}(a) using the architecture in Fig. \ref{Fig1Scheme}(c) so that the imaging system (with the exception lenses) can be monolithically integrated. In this design, a waveguide splitter allows fan-out of a single comb into multiple waveguide grating emitters and thereby multiplies pixel count and fillrate by the fan-out number $N$  (i.e., $F_N=NF_1$). The chip-based nature of microcombs can be further leveraged here to implement a monolithic microcomb array (as shown) to provide additional multiplication of the pixel count in cases where fan-out might be limited on account of comb power. For example, by using an array of one-hundred waveguide elements driven by ten, 100 GHz repetition-rate microcombs (fanout of 10:1 for each microcomb) a fillrate of 1 TeraPixels/s is possible with 10,000 pixels (see Table \ref{Design}). 100 GHz microcombs can be generated from silica microresonators with a diameter of $\sim$600 $\mu$m or silicon nitride microresonators with a diameter of $\sim$400 $\mu$m. Thus, ten microcombs may be accommodated using less than 1 cm along one linear dimension and far less along the remaining direction. On-chip grating antenna arrays are widely used for beam steering and Lidar \cite{Watts_Nature2013}. Different from beam steering, the antennas in the present application can be well separated. The spectral shower can be focused on the imaging target by two orthogonal cylindrical lenses, similar to the control of the dispersed beams in ref. \cite{Weiner_OE2008TwoD}. Then, the encoded spectral shower can be either transmitted or reflected to another receiver photonic chip with a receiver antenna array, a reference microcomb array and a photodetector array for dual-microcomb heterodyne and image retrieval. In principle, the reference microcomb array and the detector array can be integrated on the same chip with the signal microcomb array. Overall, such a design of monolithic micrcombs and grating antennas could be used to resolve transient scenes with excellent space and time resolution in the future.

As an aside, achieving wide comb bandwidth at higher repetition rates using microcombs is straightforward (e.g., 100 GHz microcombs in the 5th row design in Table \ref{Design}). However, wider bandwidth operation at reduced rates (e.g., 100 GHz microcombs in the 4th row design in Table \ref{Design}) is more challenging on account of the way continuous-wave pumping efficiency scales with repetition rate \cite{Vahala_Optica2015,Vahala_OL2016Raman}. However, pulsed pumping can be used to dramatically improve pumping efficiency and bandwidth \cite{Herr_NP2017Pulse}. For example, a 28 GHz microcomb spanning 60 THz has been demonstrated using pulse pumping \cite{Kippenberg_CLEO2019}. External (chip-based) broadeners have also be used to achieve octave-span spectral coverage for 15 GHz microcombs \cite{Diddams_PRApp2018}.

\section{Conclusion}

In conclusion, soliton microcombs have been applied to image static and moving targets, thereby validating the feasibility of using microcombs for dual-comb imaging. Further integration of microcombs with fanout waveguide grating arrays and integrated detector arrays would improve fillrate performance and image resolution. This direction of work would leverage both the spatial parallelism of conventional imaging arrays and the spectral parallelism of single-pixel optical imaging. Such systems can also eliminate discrete VIPA and grating components by bringing these functions onto the compact imaging chip. The current demonstration was in the 1.55 $\mu$m band, but can be readily shifted to the 1 $\mu$m band which could be more suitable for biological applications \cite{Vahala_NC2017Visible,Kippenberg_NC2018bio}. The method can find widespread applications in fundamental science as well as in industrial production.

%For example, it could be a useful tool to study biochemical waves in cells \cite{Petty_OPN2004} and for flow cytometry applications \cite{Jalali_PNAS2012}.

\noindent

\vspace{3 mm}
\section{Funding}
This work was supported by the Air Force Office of Scientific Research (award no. FA9550-18-1-0353), the Resnick Institute and the Kavli Nanoscience Institute at Caltech.

\section{Acknowledgments}
The authors thank Taeyoon Jeon and Chengmingyue Li for helpful discussions on the flow-cell experiment. CB gratefully acknowledges support from a Postdoctoral Fellowship from the Resnick Institute at Caltech.

\bibliography{main}
\end{document}